\documentclass[twocolumn,english,aps,prb,twocolum,superscriptaddress,bibnotes,amsmath,amssymb,floatfix]{revtex4-1}
\usepackage[colorlinks=true,citecolor=blue,linkcolor=magenta]{hyperref}
\usepackage[authormarkupposition=left]{changes} 
\usepackage{soul}
\usepackage[utf8]{inputenc}
\usepackage[english]{babel}
\usepackage{amsmath,amsfonts,amssymb}
\usepackage[T1]{fontenc}
\usepackage{url}

\usepackage{epstopdf}
\usepackage{graphicx}
\graphicspath{{./Figures/}}

\begin{document}
\title{Ultralow-loss integrated photonics enables bright, narrow-band, photon-pair sources}

\author{Ruiyang Chen}
\affiliation{Shenzhen Institute for Quantum Science and Engineering, Southern University of Science and Technology, Shenzhen 518055, China}
\affiliation{International Quantum Academy, Shenzhen 518048, China}

\author{Yi-Han Luo}
\email{luoyh@iqasz.cn}
\affiliation{International Quantum Academy, Shenzhen 518048, China}

\author{Jinbao Long}
\affiliation{International Quantum Academy, Shenzhen 518048, China}

\author{Baoqi Shi}
\affiliation{International Quantum Academy, Shenzhen 518048, China}
\affiliation{Department of Optics and Optical Engineering, University of Science and Technology of China, Hefei, Anhui 230026, China}

\author{Chen Shen}
\affiliation{International Quantum Academy, Shenzhen 518048, China}
\affiliation{Qaleido Photonics, Shenzhen 518048, China}



\author{Junqiu Liu}
\email{liujq@iqasz.cn}
\affiliation{International Quantum Academy, Shenzhen 518048, China}
\affiliation{Hefei National Laboratory, University of Science and Technology of China, Hefei 230088, China}

\maketitle

\noindent\textbf{Photon-pair sources are critical building blocks for photonic quantum systems. 
Leveraging Kerr nonlinearity and cavity-enhanced spontaneous four-wave mixing, chip-scale photon-pair sources can be created using microresonators built on photonic integrated circuit. 
For practical applications, a high microresonator quality factor $Q$ is mandatory to magnify photon-pair sources' brightness and reduce their linewidth. 
The former is proportional to $Q^4$, while the latter is inversely proportional to $Q$.  
Here, we demonstrate an integrated, microresonator-based, narrow-band photon-pair source.
The integrated microresonator, made of silicon nitride and fabricated using a standard CMOS foundry process, features ultralow loss down to $3$ dB/m and intrinsic $Q$ factor exceeding $10^7$. 
The photon-pair source has brightness of $1.17\times10^9$ Hz/mW$^2$/GHz and linewidth of $25.9$ MHz, both of which are record values for silicon-photonics-based quantum light source. 
It further enables a heralded single-photon source with heralded second-order correlation $g^{(2)}_\mathrm{h}(0)=0.0037(5)$, as well as a energy-time entanglement source with a raw visibility of $0.973(9)$. 
Our work evidences the global potential of ultralow-loss integrated photonics to create novel quantum light sources and circuits, catalyzing efficient, compact and robust interfaces to quantum communication and networks. }



Quantum science and technology have revolutionized our information society by offering a new paradigm to generate, transmit and process information. 
Photons -- travelling at $3\times10^8$ m/s and free from decoherence -- are irreplaceable carriers of quantum information \cite{Pan:12, OBrien:09}. 
Photons offer unrivalled coherence and immunity to perturbation for quantum computation \cite{WangH:19, Zhong:20, Madsen:22}, and have been ubiquitously used in quantum communication \cite{Kimble:08, Bouwmeester:1997, Lo:14, Chen:21, LuCY:22}. 
Currently, an emerging trend is to realize quantum information processing using photonic integrated circuit (PIC), which features small size, weight, and power consumption \cite{WangJW:20, Elshaari:20, Pelucchi:22}. 
In addition, PIC-based quantum chips can be manufactured with large volume and low cost using established CMOS foundries.  
Indeed, in the last decade, integrated photonics has enabled increasingly diverse applications on quantum states manipulation \cite{Vigliar:21, Bao:23}, quantum key distribution \cite{Sibson:17}, quantum networks \cite{Zheng:23}, and quantum frequency conversion \cite{Holzgrafe:20, Weaver:24}.

For photonic quantum systems, photon-pair sources are key building blocks. 
Particularly, narrow-band photon-pair sources are essential for long-distance quantum communication using quantum repeaters \cite{Gisin:07, Lvovsky:09}. 
Here the narrow bandwidth is critical, as it must match the natural linewidth of atomic transition to ensure efficient photon-atom interface. 
Meanwhile, narrow-band photon-pair sources suffer minimum distortion from chromatic dispersion when transmitting over fibers. 
Moreover, a narrow band boosts the brightness (i.e. photon flux rate) of photon-pair sources, an equally central parameter for practical applications.  

To build narrow-band photon-pair sources, nonlinear optical microresonators are particularly useful \cite{Vahala:03, Gomez:24}.
Compared to the methods using cavity-enhanced spontaneous parametric down-conversion (SPDC) with bulky crystals \cite{Ou:1999, Bao:08} or using atomic transition \cite{Balic:05, Shu:16}, optical microresonators are small, compact and robust. 
By leveraging the resonant nature for intracavity power enhancement, optical microresonators can offer exaggerated nonlinear effects, and have already been extensively used for optical frequency comb generation \cite{Kippenberg:18, Kues:19} and wideband frequency translation \cite{Li:16, Lu:19}.  

Previously, using microresonators on silicon PIC, photon-pair sources have achieved brightness of $1.6\times10^8$ Hz/mW$^2$/GHz and bandwidth of $\sim2$ GHz \cite{Ma:17}. 
The latter, due to the high optical loss (typically $>1$ dB/cm) of silicon waveguides, is overwhelmingly larger than the natural linewidth of atomic transition ($\sim 10$ MHz). 
The quest for even lower optical loss in PIC has motivated the rising and quick maturing of silicon nitride (Si$_3$N$_4$) integrated photonics. 
Integrated waveguides based on Si$_3$N$_4$ feature ultralow optical loss down to 1 dB/m \cite{Liu:21} and tailorable dispersion \cite{Okawachi:14}, and have become the leading platform for nonlinear photonics \cite{Moss:13, Gaeta:19}, narrow-linewidth lasers \cite{Jin:21, Xiang:22a}, and linear networks \cite{Taballione:19, Arrazola:21} etc. 
Indeed, Si$_3$N$_4$ has unleashed new capability for on-chip photonic quantum information processing \cite{Ramelow:15, Lu:19a, Samara:19, Samara:21, WuKY:21, Fan:23, WenW:23}. 

\begin{figure*}[t!]
\centering
\includegraphics[width=1\textwidth]{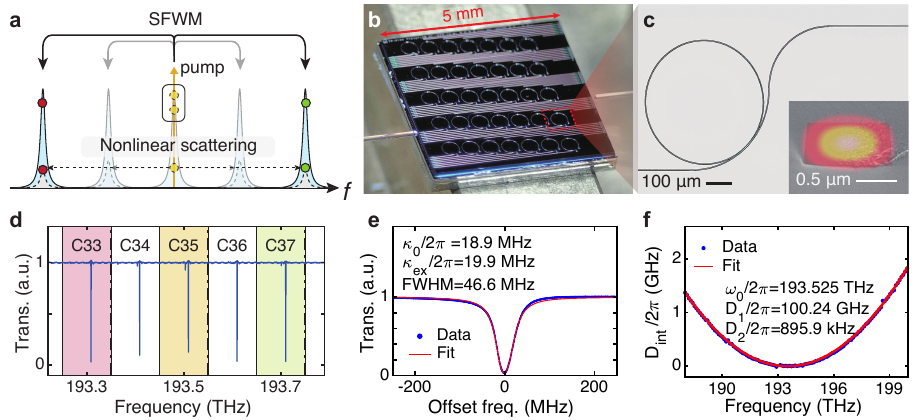}
\caption{
\textbf{Principle and characterization of integrated Si$_3$N$_4$ microresonators}. 
\textbf{a}. 
The principle of cavity-enhanced SFWM in an optical microresonator.
Two pump photons annihilate, creating a pair of signal and idler photons aligned with the resonance grid. 
Meanwhile, nonlinear scattering process can incoherently transfer pump photons to other modes. 
\textbf{b}. 
Photo of the Si$_3$N$_4$ chip containing tens of microresonators. 
Lensed fibers are used for fiber-chip light coupling. 
\textbf{c}. 
Optical microscope image showing the layout of a 100-GHz-FSR microresonator. 
The inset presents the SEM image of the Si$_3$N$_4$ waveguide's cross section, superposed with the simulated TE$_{00}$ optical mode.
\textbf{d}. 
A segment of microresonator transmission spectrum, where the pump/signal/idler mode locates in C35/C37/C33. 
\textbf{e}. 
Resonance profile in C33 with a Lorentzian fit. 
\textbf{f}. 
Measured microresonator dispersion profile fitted with Eq. \ref{eq.dint}. 
}
\label{Fig:1}
\end{figure*}
Here, we report a microresonator-based photon-pair source using ultralow-loss Si$_3$N$_4$ integrated photonics. 
The photon pairs are generated via cavity-enhanced spontaneous four-wave mixing (SFWM) in the optical microresonator \cite{Helt:10, Luo:15}, as the principle shown in Fig. \ref{Fig:1}a. 
In the presence of the microresonator's resonance grid, all the photons exist from the resonances of frequency $\omega(\mu)/2\pi$, where $\mu$ denotes the resonance mode index. 
A continuous-wave (CW) laser of frequency $\omega(\mu_\mathrm{p})/2\pi$ is used to pump the microresonator. 
Via SFWM, two pump photons annihilate, creating a pair of signal and idler photons of frequency $\omega(\mu_\mathrm{s})/2\pi$ and $\omega(\mu_\mathrm{i})/2\pi$. 
The subscript p/s/i corresponds to the pump/signal/idler mode.
In this process, the phase matching condition is automatically satisfied given $2\mu_\mathrm{p}=\mu_\mathrm{s}+\mu_\mathrm{i}$. 
The energy conservation $2\omega(\mu_\mathrm{p})=\omega(\mu_\mathrm{s})+\omega(\mu_\mathrm{i})$ requires anomalous (negative) group-velocity dispersion (GVD). 
A weak GVD is beneficial for a locally equidistant resonance grid.  
Additionally, strong optical confinement in the waveguides enhances intra-cavity electric field intensity, further facilitating SFWM. 

\noindent\textbf{Microresonator characterization}.
To satisfy these requirements, we design and fabricate integrated Si$_3$N$_4$ microresonators using a deep-ultraviolet (DUV) subtractive process on 6-inch wafers \cite{Ye:23}. 
Figure \ref{Fig:1}b shows a photograph of one final Si$_3$N$_4$ chip of $5\times5$ mm$^2$ size containing tens of microresonators. 
Figure \ref{Fig:1}c shows the microresonator's layout under an optical microscope.
The Si$_3$N$_4$ microresonators are formed with waveguides of 810-nm thickness and 2.4-$\mu$m width.
The free spectral range (FSR) of the microresonators is 100 GHz. 
Figure \ref{Fig:1}c inset shows the scanning electron microscope (SEM) image of the waveguide's cross-section, together with the simulated optical field distribution of the fundamental transverse-electric (TE$_{00}$) mode. 
It is apparent that the waveguide geometry enables tight optical confinement.
The gap between the bus waveguide and the microring resonator is 300 nm for over-coupling \cite{Cai:00}. 
A pulley-style coupler where the bus waveguide adiabatically approaches the microring is applied to ensure high coupling ideality \cite{Pfeiffer:17b}. 

\begin{figure*}[t!]
\centering
\includegraphics[width=1\textwidth]{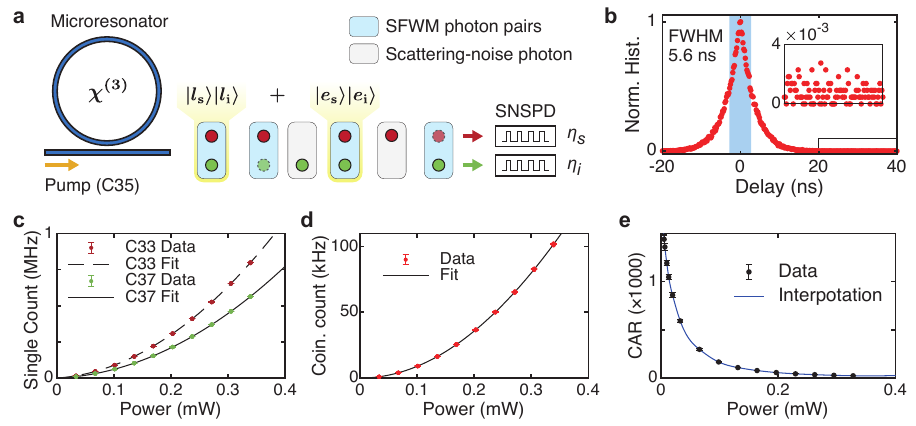}
\caption{
\textbf{Measurement of photon-pair brightness and coincidence-to-accidental ratio}.  
\textbf{a}.
Schematic of the experiment. 
The Si$_3$N$_4$ microresonator is pumped with a CW laser whose frequency locates in C35. 
The generated photons are detected with SNSPDs and analyzed with a time-to-digital converter for brightness measurement. 
The photon pair created at an earlier moment, labeled as $|e_\mathrm{i}\rangle|e_\mathrm{s}\rangle$, is coherently superposed with the photon pair $|l_\mathrm{i}\rangle|l_\mathrm{s}\rangle$ generated later, resulting in a energy-time entanglement state $|\Psi\rangle=(|e_\mathrm{i}\rangle|e_\mathrm{s}\rangle+|l_\mathrm{i}\rangle|l_\mathrm{s}\rangle)/\sqrt{2}$. 
\textbf{b}. 
The measured two-photon correlation histogram normalized to its maximum. 
The bin width is 0.2 ns. 
The histogram is obtained with on-chip pump power of $P=5.2$ $\mu$W and integration time of 3600 seconds. 
The FWHM of the peak at zero delay is 5.6 ns. 
The inset shows the zoom-in of background in the region far from the zero delay. 
\textbf{c}.
The single count rates of signal (C37, green) and idler (C33, red) photons with different on-chip power, and their polynomial fit curves. 
\textbf{d}. 
The coincident count rate with different on-chip power, and its polynomial fit. 
The coincidence window width $\Delta t = 40$ ns is chosen. 
In panels (c, d), error bars are calculated as the standard deviation of ten independent measurements, which are however smaller than the size of the data points. 
\textbf{e}. 
The measured CAR with different on-chip power.  
}
\label{Fig:2}
\end{figure*}

We characterize the Si$_3$N$_4$ microresonator using a vector spectrum analyzer in the telecommunication band \cite{Luo:24}. 
Here we study the TE$_{00}$ mode that has anomalous GVD and lowest loss among all microresonator's spatial modes. 
A segment of the microresonator's transmission spectrum is shown in Fig. \ref{Fig:1}d, where the frequency axis is calibrated with a relative precision better than 500 kHz. 
Each resonance of the 100-GHz-FSR microresonator locates in one International Telecommunication Union (ITU) channel, which can be separated using a dense wavelength-division multiplexer (DWDM). 
We plot and fit these resonances to extract the central frequency $\omega/2\pi$, the intrinsic loss $\kappa_0/2\pi$, and the external coupling strength $\kappa_\mathrm{ex}/2\pi$ \cite{Li:13}. 
As an example, the resonance profile in Channel 33 (C33) is shown in Fig. \ref{Fig:1}e, with marked $\kappa_0/2\pi$, $\kappa_\mathrm{ex}/2\pi$ and the full width at half minimum/maximum (FWHM). 
The intrinsic and loaded quality factors are calculated as $Q_0=\omega/\kappa_0=1.02\times10^7$ and $Q=\omega/\kappa=4.97\times10^6$, where $\kappa=\kappa_0 + \kappa_\mathrm{ex}$. 
The measured FWHM is larger than $\kappa/2\pi$ due to the presence of mode split \cite{Li:13} (see Note 1 in Supplementary Materials for more details). 

With the measured $\omega/2\pi$ for each resonance, the microresonator's dispersion is fitted with
\begin{equation}
  D_\mathrm{int}(\mu)=\omega(\mu)-\omega(0)-D_1\mu=D_2\mu^2/2+\mathcal{O}(\mu^3), 
\label{eq.dint}
\end{equation}
where $\omega(0)/2\pi$ is the pump resonance's frequency, 
$D_1/2\pi$ is the microresonator FSR, 
$D_2/2\pi$ is the GVD parameter. 
Higher-order dispersion terms are irrelevant in this work and thus neglected. 
The $D_\mathrm{int}$ profile and the extracted parameters are shown in Fig. \ref{Fig:1}f. 
The measured $D_2/2\pi=895.9$ kHz is sufficiently small compared to resonance linewidth $\kappa/2\pi$, satisfying energy conservation. 

\noindent\textbf{Photon-pair generation}.
Experimentally we select the mode in Channel 35 (C35) as the pump mode.  
The signal and idler photons are generated in pairs in Channel 37 (C37) and C33, as illustrated in Fig. \ref{Fig:1}d. 
The photon-pair generation rate (PGR) via cavity-enhanced SFWM is $\alpha\propto\gamma^2 P^2 Q^3$, where $\gamma$ is the effective Kerr coefficient and $P$ is the on-chip pump power \cite{Helt:10, Luo:15}. 
The photon-pair brightness is calculated as $\alpha/P^2$ divided by the photon linewidth ($\sim 0.6\kappa/2\pi$, discussed later), thus is $\propto Q^4$. 
Consequently, a high-$Q$ microresonator is extremely critical, as the photon-pair brightness is biquadratically magnified by the high $Q$. 

We note that the correct measurement of photon-pair brightness poses several requirements. 
In the ideal case that only SFWM presents, the signal and idler photons are created strictly in pairs, marked as shaded blue in Fig. \ref{Fig:2}a. 
Considering the system efficiency $\eta_\mathrm{s/i}$, single count rate for the signal/idler photon on the superconducting-nanowire single-photon detectors (SNSPD) is $n_\mathrm{s/i}=\alpha\eta_\mathrm{s/i}$, resulting in the coincidence count rate $n_\mathrm{cc}=\alpha\eta_\mathrm{i}\eta_\mathrm{s}$. 
By measuring $n_\mathrm{s}$, $n_\mathrm{i}$ and $n_\mathrm{cc}$, the PGR is calculated as 
\begin{equation}
    \alpha=n_\mathrm{s}n_\mathrm{i}/n_\mathrm{cc}
\label{eq.alpha}
\end{equation}
However, in reality, nonlinear scattering process such as Raman scattering \cite{Karpov:16} occurs in microresonators.
It can transfer the pump photons to other states, as the dashed arrows in Fig. \ref{Fig:1}a. 
The scattering-noise photons are unpaired and randomly mixed with the SFWM photon pairs, as shaded grey in Fig. \ref{Fig:2}a. 
Experimentally, scattering-noise photons cannot be distinguished from the SFWM photon pairs, as photons can be lost due to $\eta_\mathrm{s/i}<1$, marked with dashed outline in Fig. \ref{Fig:2}a. 
Therefore, Eq. \ref{eq.alpha} fails and requires modification. 

Considering that $\alpha \propto P^2$ and the generation rate of scattering-noise photons is $\alpha_\mathrm{sn} \propto P$, $n_\mathrm{s/i}$ and $n_\mathrm{cc}$ can be expressed as
\begin{align}
n_\mathrm{s/i}&=\eta_\mathrm{s/i}\left( aP^2+b_\mathrm{s/i}P \right),\\
\label{eqn:ncc}n_\mathrm{cc}-n_\mathrm{acc}&=\eta_\mathrm{i}\eta_\mathrm{s} aP^2+\mathcal{O}(P^3),
\end{align}
where $n_\mathrm{acc}$ is accidental coincidence count rate (discussed later), 
$a$ is the PGR coefficient (i.e. $\alpha=aP^2$), 
and $b_\mathrm{s/i}$ is the coefficient corresponding to the scattering-noise photon generation.
We note that, in the case that $\alpha$ and $\alpha_\mathrm{sn}$ are comparable, $n_\mathrm{acc}$ contains higher-order terms proportional to $P^3$ and $P^4$, which should be included in the righthand of Eq. \ref{eqn:ncc}. 
Since the values of $n_\mathrm{s/i}$, $n_\mathrm{cc}$ and $n_\mathrm{acc}$ can be experimentally measured at different pump power $P$, $\eta_\mathrm{s}a$, $\eta_\mathrm{i}a$ and $\eta_\mathrm{i}\eta_\mathrm{s}a$ can be extracted via polynomial fit, allowing the calculation of $a$. 
More details on the fit formula can be found in Note 2 in Supplementary Materials. 

The schematic of the experimental setup is shown in Fig. \ref{Fig:2}a, with more details found in Note 3 in Supplementary Materials. 
The photon-pair generation is evidenced by the two-photon correlation histogram shown in Fig. \ref{Fig:2}b, describing the distribution of the two-photon arrival time difference. 
The peak at the zero delay proves that the signal and idler photons are generated in pairs. 
The FWHM of the histogram peak is 5.6 ns. 
By summing up the bins within a window centered at zero delay and taking a specific width $\Delta t$, the coincidence count $n_\mathrm{cc}$ is obtained, while $n_\mathrm{acc}$ is obtained with the same $\Delta t$ but far from the zero delay. 
We emphasize that, to obtain the correct brightness, $\Delta t$ must be sufficiently large to account all the coincidence events. 
Otherwise the brightness will be overestimated.
Here we take $\Delta t=40$ ns for brightness measurement. 
More details can be found in Note 4 in Supplementary Materials. 

The measured $n_\mathrm{s/i}$, $n_\mathrm{cc}$ and $n_\mathrm{acc}$ versus on-chip power $P$, together with their fit curves, are shown in Figs. \ref{Fig:2}d and \ref{Fig:2}e, respectively.
The error bars are calculated as the standard deviation of ten independent measurements.
Experimentally we obtain $a=3.04\times10^7$ Hz/mW$^2$. 
Detailed fit results are found in Note 4 in Supplementary Materials. 
The generated photons' linewidth is estimated as $\delta\nu=25.9$ MHz. 
The brightness is thus calculated as $1.17\times10^9$ Hz/mW$^2$/GHz. 
More details on the linewidth estimation are elaborated in Note 5 in Supplementary Materials.

In addition to the brightness, the noise level is equally important.
As shown in Fig. \ref{Fig:2}b, for the region far outside the zero delay, the bin counts are not exactly zero, mainly due to the SFWM multi-photon excitation and the scattering noise. 
The coincidence-to-accidental ratio (CAR) is defined as $(n_\mathrm{cc}/n_\mathrm{acc}-1)$. 
Figure \ref{Fig:2}c shows the measured CAR with $\Delta t=5.6$ ns, i.e. the FWHM of the histogram peak in Fig. \ref{Fig:2}b. 
As the pump power $P$ decreases, both the SFWM multi-photon excitation and the scattering noise decrease, resulting in a higher CAR.
Experimentally, with $P=5.2$ $\mu$W, we achieve $n_\mathrm{cc}/n_\mathrm{acc}-1=1438\pm22$ with 3600 second integration. 
Meanwhile, the measured coincidence count rate is $n_\mathrm{cc}=22$ Hz.

\begin{figure}[t!]
\centering
\includegraphics[width=0.5\textwidth]{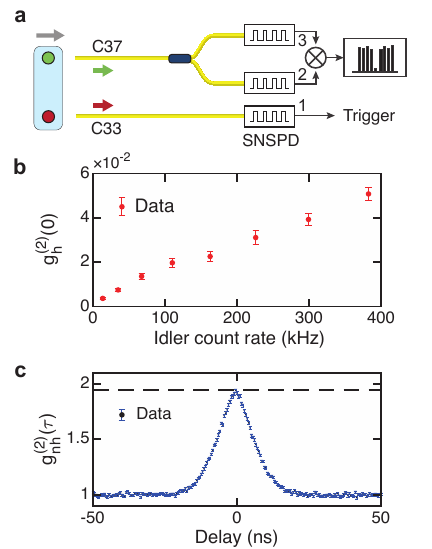}
\caption{
\textbf{Measurement of the second-order correlation}. 
\textbf{a}. 
Schematic of the experimental setup. 
The idler photons (C33, red) are directly detected with an SNSPD as trigger (marked as path 1). 
The signal photons (C37, green) are equally splitted and detected (marked as paths 2 and 3). 
The two-photon correlation histogram between paths 2 and 3 is measured with and without the path 1 clicking. 
\textbf{b}. 
Measured $g^{(2)}_\mathrm{h}(0)$ versus different idler photon count rate. 
\textbf{c}. 
Measured $g^{(2)}_\mathrm{nh}(\tau)$. 
The bin width is chosen as 0.5 ns. 
The maximum $g^{(2)}_\mathrm{nh}(\tau)=1.942(14)$ is achieved at zero delay. 
}
\label{Fig:3}
\end{figure}

\begin{figure*}[t!]
\centering
\includegraphics[width=1\textwidth]{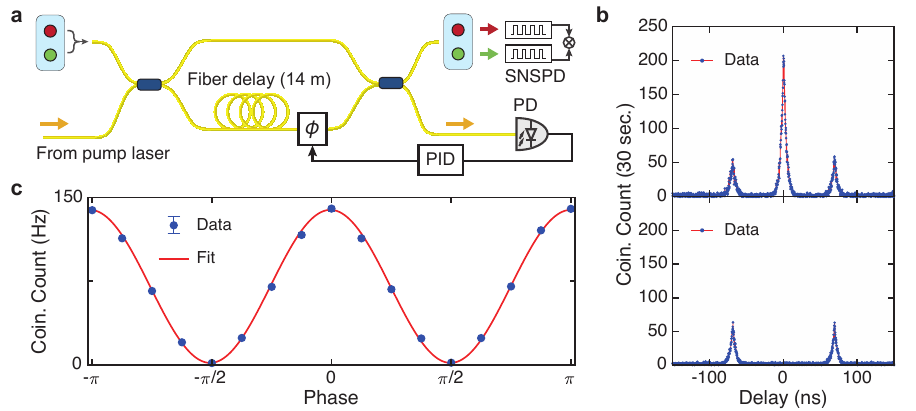}
\caption{
\textbf{Two-photon interference and energy-time entanglement}. 
\textbf{a}. 
Schematic of the experimental setup. 
To avoid single-photon interference, a 14-meter-long fiber link is used to introduce 69 ns delay. 
A portion of the pump laser is used to actively stabilize the phase difference between the two branches via controlling the added phase $\phi$ caused by the fiber stretcher. 
\textbf{b}. 
Two-photon correlation histogram. 
In the upper/lower panel, the peak at zero delay appears/vanishes due to constructive/destructive two-photon interference. 
\textbf{c}. 
Two-photon interference fringe. 
The sinusoidal fit shows a raw visibility of $V=0.973(9)$ without background subtraction. 
Error bars are much smaller than the point size and thus are invisible.
}
\label{Fig:4}
\end{figure*}

\noindent\textbf{Second-order correlation}.
Photon-pair sources can be used as heralded single-photon sources.
Once the SNSPD clicks upon the arrival of an idler photon, it heralds the existence of a signal photon. 
The photon anti-bunching can be observed with the heralded signal photons. 
To characterize the single-photon purity, a Hanbury Brown and Twiss (HBT) setup is utilized to measure the heralded second-order correlation $g^{(2)}_\mathrm{h}(0)$. 
As shown in Fig. \ref{Fig:3}a, the DWDM's C37 is equally splitted into two branches with marked path indices 2 and 3. 
Together with C33 of path index 1, the $g^{(2)}_\mathrm{h}(0)$ is calculated as $g^{(2)}_\mathrm{h}(0) = n_{123}n_1/(n_{12}n_{13}$), 
where $n_1$ is the single count rate,  $n_{12}$, $n_{13}$, $n_{123}$ are the coincidence count rates. 
Experimentally, we observe $g^{(2)}_\mathrm{h}(0) = 0.0037(5)$ at idler count rate $n_\mathrm{s}=13.3$ kHz, with $P=40$ $\mu$W and $\Delta t=5.6$ ns.
Due to the SFWM multi-pair excitation, $g^{(2)}_\mathrm{h}(0)$ increases with increasing $P$ (thus $n_\mathrm{i}$), as shown in Fig. \ref{Fig:3}b. 

Next we characterize the spectral purity of the photon pairs by measuring the non-heralded second-order correlation $g^{(2)}_\mathrm{nh}(\tau)$ \cite{Christ:11}. 
Such measurement does not involve idler photons. 
The $g^{(2)}_\mathrm{nh}(\tau)$ histogram between paths 2 and 3 is shown in Fig. \ref{Fig:3}c. 
It is observed that $g^{(2)}_\mathrm{nh}(\tau)=1.942(14)$ is reached at zero delay. 
It indicates that, a high multi-photon interference visibility between two independent photon-pair sources requires a narrow coincidence window \cite{Huang:10}. 
This is mainly due to the strong spectral correlation of the photon pairs generated with the CW pump. 
The influence of the spectral correlation can be eliminated by introducing a dual interference scheme \cite{LiuRZ:23}. 

\noindent \textbf{Two-photon interference visibility}.
Photon-pair sources can be regarded as narrow-band, energy-time entanglement light sources, which are advantageous for long-distance distribution due to their robustness against decoherence in optical fibers \cite{Marcikic:04, Inagaki:13, YuY:20}.
As shown in Fig. \ref{Fig:2}a, the photon pair created at an earlier moment, labeled as $|e_\mathrm{i}\rangle|e_\mathrm{s}\rangle$, is coherently superposed with the photon pair $|l_\mathrm{i}\rangle|l_\mathrm{s}\rangle$ generated later, where $e$/$l$ denotes the earlier/later generated photon and the subscript s/i marks the signal/idler photon. 
Thus energy-time entangled photons $|\Psi\rangle=(|e_\mathrm{i}\rangle|e_\mathrm{s}\rangle+|l_\mathrm{i}\rangle|l_\mathrm{s}\rangle)/\sqrt{2}$ are created. 
To reveal the quantum interference, the two photons are projected along the state $|\phi\rangle=(|e\rangle+e^{i\phi}|l\rangle)/\sqrt{2}$ individually. 
The probability of measuring the state $|\Psi\rangle$ in $|\phi_\mathrm{i}\rangle|\phi_\mathrm{s}\rangle$ is calculated as $p=(1+\cos 2\phi)/4$. 

We use a folded Franson interferometer \cite{Franson:89, Ou:90} to interfere two temporally separated photons. 
The setup is shown in Fig. \ref{Fig:4}a. 
The photons are equally splitted into two branches. 
In the lower branch, a 14-meter-long fiber link, corresponding to 69 ns delay, is introduced to avoid single-photon interference. 
A fiber stretcher is used to control the added phase $\phi$. 
The two branches are then recombined and interfere. 
To suppress high-frequency phase fluctuation in the fiber due to ambient temperature fluctuation, we place the interferometer in a heat-insulated container. 
Additionally, a portion of the pump laser is used for
active phase locking with a PID of 10 kHz bandwidth. 
The phase $\phi$ is varied by tuning the locking voltage and the PID's locking point. 

The two-photon interference is evidenced by the two-photon correlation histogram shown in Fig. \ref{Fig:4}b. 
The central peak is post-selected, as it corresponds to the measurement of $|\Psi\rangle$. 
For $\phi=0$, the central peak reaches maximum that is fourfold to the sidebands. 
When $\phi=\pi/2$, the central peak vanishes due to destructive interference. 
Detailed analysis is found in Note 6 in Supplementary Materials. 

The measured two-photon interference is shown in Fig. \ref{Fig:4}c, where $\Delta t=5.6$ ns is identical to that of CAR measurement. 
The coincidence count $n_\mathrm{cc}$ oscillates with a period of $\pi$.
A sinusoidal fit using $n_\mathrm{cc}=0.5N[1+V\cos(\phi)]$ is applied to extract a raw visibility $V=0.973(9)$ at $P=97$ $\mu$W pump power, well above 0.707 and thus violating Clauser–Horne–Shimony–Holt (CHSH) inequality \cite{Clauser:69}. 
Due to the scattering noise, the background of the histogram in Fig. \ref{Fig:4}b is much larger than that of a crystal-based photon-pair source \cite{Zhong:18}. 
The background can severely deteriorate the interference visibility. 
In our experiment, the background is calculated by averaging the bin values far from the peaks. 
By subtracting the averaged background, we obtain a visibility of $V=0.995(9)$. 

\noindent\textbf{Conclusion and discussion}.
In summary, we have studied integrated, high-$Q$, Kerr-nonlinear microresonators for bright, narrow-band, photon-pair generation.
Using an over-coupled microresonator on ultralow-loss Si$_3$N$_4$ PIC and fabricated with a standard CMOS-foundry process, we realize a photon-pair source with brightness exceeding $1.17\times10^9$ Hz/mW$^2$/GHz and bandwidth below 25.9 MHz, both of which are record values for silicon-photonics-based quantum light sources. 
Moreover we achieve a CAR of $n_\mathrm{cc}/n_\mathrm{acc}-1=1438\pm22$, a heralded second-order correlation of $g^{(2)}_\mathrm{h}(0)=0.0037(5)$, and a raw two-photon interference visibility of $V=0.973(9)$. 
A comparison of integrated quantum light sources using microresonators is presented in Note 7 in Supplementary Materials. 

Since the optical loss of our Si$_3$N$_4$ waveguides is far above the material limit of absorption loss \cite{Liu:21, Gao:22}, there is major space to further improve the microresonator $Q$ factors by optimizing the material growth and fabrication process. 
Consequently, the brightness ($\propto Q^4$) can well exceed the best performance of recently developed highly nonlinear AlGaAs-on-insulator microresonators \cite{Steiner:21, Chang:20}.
Combined with the wide transparency window of Si$_3$N$_4$ and broadband dispersion engineering, photon-pair generation from the visible to mid-infrared can be envisaged, paving the way to interfacing integrated photonics with a variety of quantum devices.

\medskip
\begin{footnotesize}

\noindent \textbf{Acknowledgments}: 
We thank Yuan Cao and Hui-Nan Wu for the fruitful discussion on Franson interferometer phase locking, Yun-Ru Fan for the suggestion on the energy-time entanglement measurement, and Shuyi Li for the suggestion on second-order correlation measurement.
J. Liu acknowledges support from the National Natural Science Foundation of China (Grant No.12261131503), Innovation Program for Quantum Science and Technology (2023ZD0301500), Shenzhen-Hong Kong Cooperation Zone for Technology and Innovation (HZQB-KCZYB2020050), and from the Guangdong Provincial Key Laboratory (2019B121203002).
Y.-H L. acknowledges support from the China Postdoctoral Science Foundation (Grant No. 2022M721482). 
The silicon nitride chips were fabricated by Qaleido Photonics. 
R. C. and Y.-H. L. contributed equally to this work. 

\noindent \textbf{Author contributions}: 
Y.-H. L. and J. Liu conceived the experiment. 
R. C. and Y.-H. L. built the experimental setup, assisted with J. Long. 
S. C. fabricated the Si$_3$N$_4$ chip device. 
B. S. characterized the Si$_3$N$_4$ chip device. 
R. C., Y.-H. L. and J. Liu analyzed the data and prepared the manuscript with input from others. 
J. Liu supervised the project.  

\noindent \textbf{Conflict of interest}:
The authors declare no conflicts of interest. 

\noindent \textbf{Data Availability Statement}: 
The code and data used to produce the plots within this work will be released on the repository \texttt{Zenodo} upon publication of this preprint.

\end{footnotesize}

\bibliographystyle{apsrev4-1}
%

\end{document}


\title{Supplementary Materials for: Ultralow-loss integrated photonics enables bright, narrow-band, photon-pair sources}

\author{Ruiyang Chen}
\affiliation{Shenzhen Institute for Quantum Science and Engineering, Southern University of Science and Technology, Shenzhen 518055, China}
\affiliation{International Quantum Academy, Shenzhen 518048, China}

\author{Yi-Han Luo}
\email{luoyh@iqasz.cn}
\affiliation{International Quantum Academy, Shenzhen 518048, China}

\author{Jinbao Long}
\affiliation{International Quantum Academy, Shenzhen 518048, China}

\author{Baoqi Shi}
\affiliation{International Quantum Academy, Shenzhen 518048, China}
\affiliation{Department of Optics and Optical Engineering, University of Science and Technology of China, Hefei, Anhui 230026, China}

\author{Chen Shen}
\affiliation{International Quantum Academy, Shenzhen 518048, China}
\affiliation{Qaleido Photonics, Shenzhen 518048, China}

\author{Junqiu Liu}
\email{liujq@iqasz.cn}
\affiliation{International Quantum Academy, Shenzhen 518048, China}
\affiliation{Hefei National Laboratory, University of Science and Technology of China, Hefei 230088, China}

\maketitle

\section{Characterization and fitting of microresonator resonances}
\vspace{0.5cm}

\begin{figure*}[h!]
\centering
\includegraphics[width=0.95\textwidth]{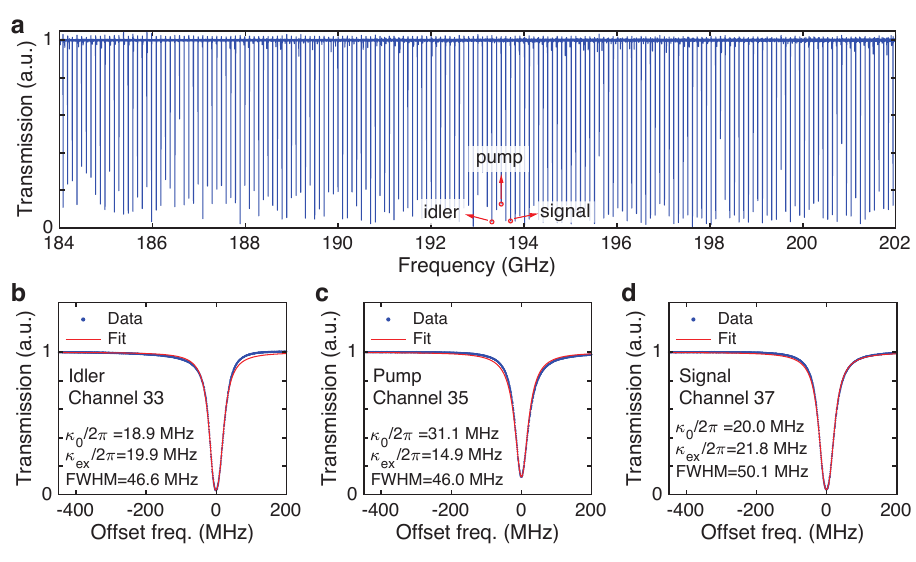}
\caption{
\textbf{Microresonator transmission spectrum and resonances}. 
\textbf{a}. 
Transmission spectrum of the Si$_3$N$_4$ microresonator ranging from 184 THz to 202 THz. 
The pump, signal and idler modes are marked with red circles. 
\textbf{b, c, d}.
Resonance profiles of the Channels 33, 35 and 37, together with their fit curves. 
The intrinsic loss $\kappa_0/2\pi$, the external coupling rate $\kappa_\mathrm{ex}/2\pi$ and the FWHM of each resonance are marked. 
}
\label{Fig:S1}
\end{figure*}

We use a tunable diode laser that sweeps from 184 THz (1640 nm) to 202 THz (1480 nm) to characterize the resonances of the Si$_3$N$_4$ microresonator \cite{Luo:24}. 
The light exiting the microresonator is detected with a photodiode, and the transmission is recorded with an oscilloscope. 
First, the raw microresonator transmission spectrum is normalized to the non-resonant region, as shown in Fig. \ref{Fig:S1}a. 
The resonant frequencies are extracted via dip searching, and later used for dispersion fit, as shown in Fig. 1f in the main text. 
The resonances corresponding to the pump, signal, and idler modes are marked with red circles in Fig. \ref{Fig:S1}a. 
For each resonance, we fit the profile using the formula \cite{Li:13}
\begin{equation}
    T = \left|1 - \frac{\kappa_\mathrm{ex}\left[i(\omega-\omega_0)+(\kappa_0+\kappa_\mathrm{ex})/2\right]}{\left[i(\omega-\omega_0)+(\kappa_0+\kappa_\mathrm{ex})/2\right]^2+(k_\mathrm{r}+ik_\mathrm{i})^2/4} \right|^2, 
    \label{eqn:S1}
\end{equation}
where $k_\mathrm{r}$/$k_\mathrm{i}$ is the real/imaginary part of the complex coupling coefficient of clockwise and counter-clockwise modes, in the presence of mode split. 
As $\kappa_0$ and $\kappa_\mathrm{ex}$ are symmetric in Eq. (\ref{eqn:S1}), it is impossible to distinguish $\kappa_0$ and $\kappa_\mathrm{ex}$ simply via the fit. 
Considering that in Fig. \ref{Fig:S1}a, the resonance dips become deeper as the tunable laser sweeps from 184 THz to $\sim$198 THz, we infer that the microresonator is over-coupled in this region, i.e. $\kappa_\mathrm{ex}>\kappa_0$. 
For resonances in the region of $\sim$198 THz to 202 THz, we treat $\kappa_\mathrm{ex}<\kappa_0$. 
This can also be verified with phase response measurement \cite{Luo:24}. 
Therefore $\kappa_0$ and $\kappa_\mathrm{ex}$ for each resonance can be extracted. 
The resonances corresponding to C33 (idler), C35 (pump) and C37 (signal), together with their fit curves, are shown in Fig. \ref{Fig:S1}(b, c, d). 
The extracted $\kappa_0/2\pi$, $\kappa_\mathrm{ex}/2\pi$ and the full width at half minimum/maximum (FWHM) are marked in the figures.
It should be noted that, due to the mode split, i.e. coupling between clockwise and counter-clockwise modes, the FWHM is larger than $(\kappa_0 + \kappa_\mathrm{ex})/2\pi$. 

\clearpage
\section{Fit formula for brightness measurement}
\vspace{0.5cm}

In the experiment, photons in the signal and idler modes enter two independent SNSPDs, generating two series of photon clicking events. 
Each event has a specific time tag, thus can be unambiguously placed on the time axis. 
The time axis is uniformly divided into segments with a width of $\delta t=1$ second/$M$. 
The integer $M$ is chosen such that $\delta t$ is comparable to the coherence time of the two-photon wave package. 
If the clicking events of a signal photon and an idler photon are found in the same segment, these two photons are considered to be generated in pairs, resulting in a coincidence event. 

Besides SFWM, photons can also be generated from nonlinear scattering processes that cause accidental coincidence events. 
To accurately determine the photon-pair brightness, we must subtract the contribution of nonlinear scattering processes.
Considering that the photon-pair generation rate via SFWM is proportional to $P^2$, and the generation rate of scattering-noise photons is proportional to $P$, one can distinguish these two types of photons via polynomial fit of the coincidence count rate at different pump power $P$. 
To obtain the exact fit formula, the origins of the coincidence count rate $n_\mathrm{cc}$ and the accidental coincidence count rate $n_\mathrm{acc}$ must be carefully analyzed.
In the following, the definition of each variable is consistent with the main text. 

\vspace{1mm}
The experimentally measured $n_\mathrm{cc}$ consists of three parts:
\begin{itemize}
\item The photon pairs generated via the SFWM process naturally contribute a coincidence count rate of $\eta_\mathrm{s}\eta_\mathrm{i}aP^2$. 

\item The single-photon events in the signal and idler modes resulting from the nonlinear scattering processes accidentally coincide, leading to accidental coincidence events.
Considering the scattering-noise photon generation rate $b_\mathrm{s/i}P$, for each segment, there exists a noise photon with a probability of $p_\mathrm{s/i} = b_\mathrm{s/i}P/M$. 
Therefore, the corresponding coincidence count rate on the SNSPD can be calculated as $\eta_\mathrm{s}\eta_\mathrm{i}\cdot Mp_\mathrm{s}p_\mathrm{i}=\eta_\mathrm{s}\eta_\mathrm{i}b_\mathrm{s}b_\mathrm{i}P^2/M$. 

\item Due to the non-ideal photon detection efficiency $\eta<1$ ($\eta\sim 0.2$ in our experiment), an SFWM-generated photon may miss its partner but meet a scattering-noise photon in the other mode, contributing an accidental coincidence term scaling as $\mathcal{O}(P^3)$ in $n_\mathrm{cc}$.
\end{itemize}

Therefore, we have
\begin{equation}
    n_\mathrm{cc} = \eta_\mathrm{s}\eta_\mathrm{i}aP^2 + \eta_\mathrm{s}\eta_\mathrm{i}b_\mathrm{s}b_\mathrm{i}P^2/M + \mathcal{O}(P^3). 
    \label{eq:ncc}
\end{equation}

From the above analysis, one can see that the measured $n_\mathrm{cc}$ inevitably includes accidental coincidence events. 
To extract the pure contribution of the SFWM process, we also measure the accidental coincidence count rate $n_\mathrm{acc}$. 
In the experiment, $n_\mathrm{acc}$ is measured using the same method as for $n_\mathrm{cc}$, but with a large delay between the signal and the idler modes. 
Thus all the coincidence events occur randomly. 
Similarly, a single temporal segment for the signal/idler mode contains a photon with a probability of $p_\mathrm{s/i}=\eta_\mathrm{s/i}(aP^2 + b_\mathrm{s/i}P)/M$, thus $n_\mathrm{acc}$ can be calculated as
\begin{equation}
    n_\mathrm{acc} = Mp_\mathrm{s}p_\mathrm{i}=\eta_\mathrm{s}\eta_\mathrm{i}b_\mathrm{s}b_\mathrm{i}P^2/M +\mathcal{O}(P^3). 
    \label{eq:nacc}
\end{equation}

\begin{figure*}[h!]
\centering
\includegraphics[width=0.88\textwidth]{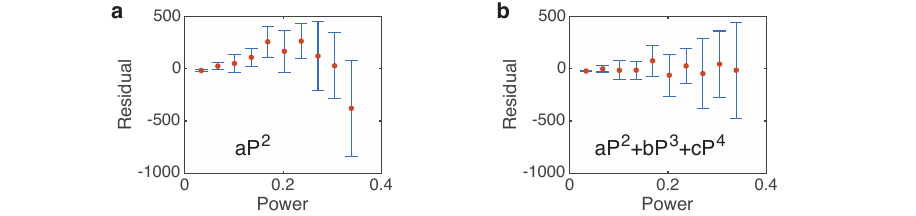}
\caption{
\textbf{Fit residual of the brightness measurement}. 
\textbf{a} and \textbf{b} correspond to the polynomial fit without and with the $P^3$ and $P^4$ terms. 
A clear trend is seen in \textbf{a}, indicating necessity to include $P^3$ and $P^4$ terms. 
}
\label{Fig:Res}
\end{figure*}

Combining Eq. (\ref{eq:ncc}) and (\ref{eq:nacc}), we have $n_\mathrm{cc}-n_\mathrm{acc}=\eta_\mathrm{i}\eta_\mathrm{s}aP^2+\mathcal{O}(P^3)$, which is Eq. (2) in the main text.
Experimentally, we fit $(n_\mathrm{cc}-n_\mathrm{acc})$ with a polynomial containing $P^2$, $P^3$, and $P^4$ terms. 
The coefficient of the $P^2$ term is $\eta_\mathrm{i}\eta_\mathrm{s}a$. 
With $\eta_\mathrm{s}a$ and $\eta_\mathrm{i}a$ extracted using the single count rates' fit, the coefficient $a$ can be determined. 
To experimentally verify the necessity of the high-order terms ($P^3$ and $P^4$), the fit residual (i.e. data deviations from the fit curve) without and with the high-order terms are compared in Fig. \ref{Fig:Res}a and \ref{Fig:Res}b. 

\clearpage
\section{Experiment setup}
\vspace{0.5cm}

\begin{figure*}[h!]
\centering
\includegraphics[width=0.95\textwidth]{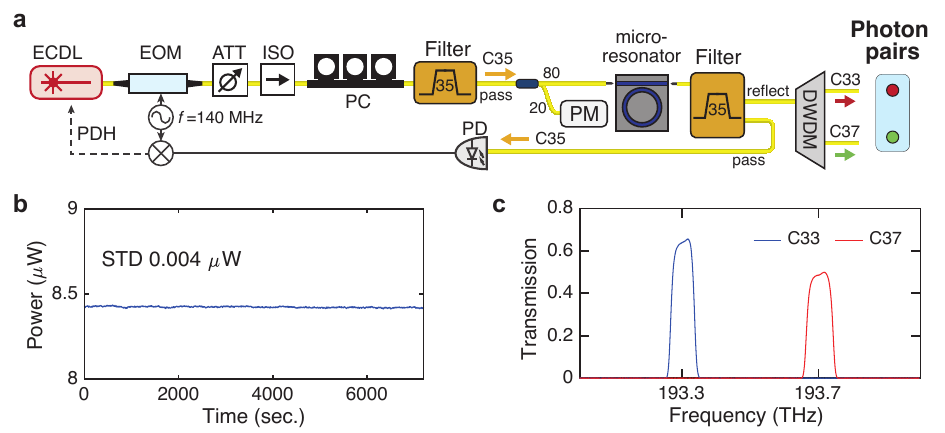}
\caption{
\textbf{Experiment setup}. 
\textbf{a}. 
EOM, electro-optic modulator;
ATT, optical attenuator;
ISO, optical isolator;
PM, powermeter;
PC, polarization controller;
PD, photodetector;
\textbf{b}.
Measured stability of the fiber-chip coupling efficiency. 
A powermeter is used to monitor the output power from the chip over 7200 seconds. 
The standard deviation is 0.004 $\mu$W. 
\textbf{c}. 
The overall transmission profiles of the fiber link between the output lensed fiber and the photon detector for C33 and C37. 
}
\label{Fig:S2}
\end{figure*}

The experimental setup to generate photon pairs is shown in Fig. \ref{Fig:S2}a. 
The pump laser is an external-cavity diode laser (ECDL) PDH-locked to the resonant frequency of the pump mode. 
An optical attenuator is used to vary the pump power, monitored by a powermeter. 
A 100-dB bandpass filter is used to reject the laser's spontaneous emission outside of C35. 
The pump laser is then edge-coupled into and out of the microresonator using lensed fibers \cite{Liu:18}. 

The fiber-chip coupling efficiency is calibrated and stabilized, critical for calculating the on-chip pump power $P$.  
We experimentally verify the coupling stability as shown in Fig. \ref{Fig:S2}b. 
The out-coupled light from the chip is monitored with a powermeter over 7200 seconds. 
During this period, the power is averaged as 8.423(4) $\mu$W.
The exceptional coupling stability ensures the reliability and reproducibility of the brightness measurement. 

The on-chip pump power $P$ is evaluated as following. 
Before sending the light into the chip, the pump light in the input lensed fiber is splitted with a 20:80 beam splitter. 
The 80\% branch is coupled into the chip, while the 20\% branch is measured as $P_\mathrm{m}$ by a powermeter. 
Thus $P=4P_\mathrm{m}\cdot\eta_\mathrm{cp}\eta_\mathrm{f}$, where $\eta_\mathrm{cp}$ is the the fiber-chip coupling efficiency, $\eta_\mathrm{f}=0.96$ is the transmission of the flange before the lensed fiber.
In the experiment, we measure the two-sided coupling efficiency ($\eta_\mathrm{cp}^2$) of twenty microresonator devices of identical design on several chips with the same lensed fibers, which is averaged as $\eta_\mathrm{cp}^2=0.467(12)$. 
For the microresonator used in this work, $\eta_\mathrm{cp}^2=0.471$, consistent with the statistical result. 
Therefore, we evaluate the on-chip power $P=2.635P_\mathrm{m}$. 

In the experiment, the output signal and idler photons from the microresonator are reflected by another 100-dB bandpass filter and separated with a dense wavelength-division multiplexer (DWDM).
Then, the photons are directed to SNSPDs, or to the setups (shown in Fig. 3a and 4a in the main text) for the second-order correlation measurement and the energy-time entanglement state preparation. 
Here, we measure the overall efficiency of the filter along with the DWDMs for C33 and C37, as shown in Fig. \ref{Fig:S2}c. 
The transmission coefficients corresponding to the passband center of C37 (signal) and C33 (idler) are 0.488 and 0.642, respectively.  

\clearpage
\section{Brightness measurement}
\vspace{0.5cm}

\begin{figure*}[h!]
\centering
\includegraphics[width=0.35\textwidth]{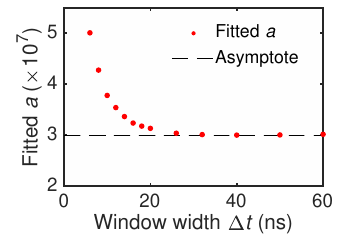}
\caption{
\textbf{The fitted $\bm{a}$ describing brightness versus the coincidence window width $\bm{\Delta t}$}. 
As $\Delta t$ increases, the fitted $a$ decreases and approaches a constant value.
We choose $\Delta t=40$ ns in the experiment to measure the brightness. 
}
\label{Fig:S3}
\end{figure*}

For the brightness measurement, the selection of the temporal window width $\Delta t$ is crucial for the coincidence count rate $n_\mathrm{cc}$. 
A $\Delta t$ value smaller than the two-photon wave package leads to additional loss for two-photon events, which is equivalent to modifying Eq. 2 in the main text as
\begin{equation}
    n_\mathrm{cc} - n_\mathrm{acc} = \eta_\mathrm{win}(\Delta t)\cdot\eta_\mathrm{i}\eta_\mathrm{s}aP^2 + \mathcal{O}(P^3). 
\end{equation}
Experimentally, the loss $\eta_\mathrm{win}(\Delta t) < 1$ is attributed to the overall system loss, making the extracted $\eta_\mathrm{s}\eta_\mathrm{i}a$ to be $\eta_\mathrm{win}(\Delta t)\eta_\mathrm{s}\eta_\mathrm{i}a$. 
Note that the single count rate is irrelevant to $\Delta t$, thus underestimating $\eta_\mathrm{i}\eta_\mathrm{s} a$ leads to an overestimation of photon-pair generation rate $\alpha$. 
Therefore, for the brightness measurement, $\Delta t$ should be wide enough to account for all the coincidence events resulted from SFWM. 
We verify this condition experimentally. 
Figure \ref{Fig:S3} shows that, as $\Delta t$ increases, the measured brightness decreases and approaches a constant value. 
In our experiment, $\Delta t=40$ ns is selected. 

We perform polynomial fit on the single count rate $n_\mathrm{s/i}(P)$ and the coincidence count rate $n_\mathrm{cc/acc}(P)$ with different pump power $P$. 
The result is following: 
$\eta_\mathrm{i}a = 6.137\times10^6$ Hz/mW$^2$,
$\eta_\mathrm{s}a = 4.419\times10^6$ Hz/mW$^2$, 
$\eta_\mathrm{s}\eta_\mathrm{i}a = 8.907\times10^5$ Hz/mW$^2$. 
The overall efficiency values for the signal (C37) and idler (C33) modes are calculated as $\eta_\mathrm{s}=0.145$ and $\eta_\mathrm{i}=0.202$. 
Considering the fiber-chip coupling efficiency $\eta_\mathrm{cp}=0.683$ and the averaged detection efficiency of the SNSPD of 0.830, the coupling efficiency values between the Si$_3$N$_4$ microresonator and the bus waveguide are extracted to be 0.524 and 0.555 for the signal and idler modes, which are consistent with the results calculated with $\kappa_\mathrm{ex}/\kappa$. 

\clearpage
\section{Linewidth of the SFWM generated photons}
\vspace{0.5cm}

Experimentally the intrinsic loss $\kappa_0/2\pi$ and the external coupling rate $\kappa_\mathrm{ex}/2\pi$ are extracted. 
However, the resonance linewidth $\kappa/2\pi=(\kappa_0+\kappa_\mathrm{ex})/2\pi$ does not exactly equal to the linewidth of the generated photons. 
For example, for cavity-enhanced spontaneous parametric down-conversion (SPDC), the generated photons have linewidth \cite{Ou:1999} of $0.64\cdot\kappa/2\pi$. 
Typically, different resonances of the same optical microresonator have different $\kappa/2\pi$ values. 
Here, we address the photon linewidth in cavity-enhanced SFWM. 

In our experiment, a pump beam enters the microresonator and excites photon pairs in signal (s) and idler (i) modes, respectively. 
The system's Hamiltonian can be written as
\begin{equation}
H_\mathrm{sys} = \hbar(\omega_\mathrm{0, p} + \Delta) a_\mathrm{i}^\dagger a_\mathrm{i} + \hbar(\omega_\mathrm{0, p} - \Delta) a_\mathrm{s}^\dagger a_\mathrm{s} + i\hbar\gamma\left[
	\epsilon^2 e^{-i2\omega_\mathrm{p} t }a_\mathrm{i}^\dagger a_\mathrm{s}^\dagger - \epsilon^{*2} e^{i2\omega_\mathrm{p} t }a_\mathrm{i} a_\mathrm{s}
\right], 
\label{eqn:hamiltonian}
\end{equation}
where $\omega_\mathrm{0, p}$ is the resonant frequency corresponding to the pump mode, 
$\Delta$ is the microresonator FSR, 
$\omega_\mathrm{p}$ is the frequency of the pump beam, 
$\gamma$ is the effective Kerr nonlinear coefficient, 
$\epsilon$ is the intra-cavity amplitude of the pump beam. 
The Hamiltonian is then inserted into the Langevin equation
\begin{equation}
\frac{\mathrm{d}a_\mathrm{i/s}}{\mathrm{d}t} = -\frac{i}{\hbar}[a_\mathrm{i/s}, H_\mathrm{sys}] - \frac{\kappa_\mathrm{i/s}}{2} a_\mathrm{i/s} + \sqrt{\kappa_\mathrm{i/s, \mathrm{ex}}} a_\mathrm{i/s, \mathrm{in}} + \sqrt{\kappa_\mathrm{i/s, 0}} a_\mathrm{i/s, \mathrm{loss}}, 
\end{equation}
where $\kappa_\mathrm{i/s} = \kappa_\mathrm{i/s, \mathrm{0}} + \kappa_\mathrm{i/s, \mathrm{ex}}$ is the total loss rate, 
$\kappa_\mathrm{i/s, \mathrm{0}}$ is the intrinsic loss rate, 
$\kappa_\mathrm{i/s, \mathrm{ex}}$ is the external coupling rate, 
and the subscript i/s denote the idler/signal mode. 
By taking the transformation $a_\mathrm{i}\rightarrow e^{-i(\omega_\mathrm{0, p} + \Delta)t}a_\mathrm{i}$ and $a_\mathrm{s}\rightarrow e^{-i(\omega_\mathrm{0, p} - \Delta)t}a_\mathrm{s}$, the Langevin equations of the system become
\begin{subequations}
\begin{align}
\frac{\mathrm{d}a_\mathrm{i}}{\mathrm{d}t} &= \gamma\epsilon^2 a_\mathrm{s}^\dagger - \frac{\kappa_\mathrm{i}}{2}a_\mathrm{i} + \sqrt{\kappa_\mathrm{i, \mathrm{ex}}} a_\mathrm{i, \mathrm{in}} + \sqrt{\kappa_\mathrm{i, 0}} a_\mathrm{i, \mathrm{loss}}, \\[4pt]
\frac{\mathrm{d}a_\mathrm{s}}{\mathrm{d}t} &= \gamma\epsilon^2 a_\mathrm{i}^\dagger - \frac{\kappa_\mathrm{s}}{2}a_\mathrm{s} + \sqrt{\kappa_\mathrm{s, \mathrm{ex}}} a_\mathrm{s, \mathrm{in}} + \sqrt{\kappa_\mathrm{s, 0}} a_\mathrm{s, \mathrm{loss}}.
\end{align}
\label{eq:final}
\end{subequations}

Equations (\ref{eq:final}) are then transformed to the frequency domain, solved for the steady state of the system
\begin{subequations}
\begin{align}
a_\mathrm{i}(\omega) = \frac{\sqrt{\kappa_\mathrm{i, \mathrm{ex}}}(\kappa_\mathrm{s}/2 - i\omega) a_\mathrm{i, \mathrm{in}} + \sqrt{\kappa_\mathrm{i, \mathrm{0}}}(\kappa_\mathrm{s}/2 - i\omega) a_\mathrm{i, \mathrm{loss}} + \gamma\epsilon^2\sqrt{\kappa_\mathrm{s, \mathrm{ex}}} a_\mathrm{s, \mathrm{in}}^\dagger + \gamma\epsilon^2\sqrt{\kappa_\mathrm{s, \mathrm{0}}} a_\mathrm{s, \mathrm{loss}}^\dagger}{(\kappa_\mathrm{i}/2-i\omega)(\kappa_\mathrm{s}/2-i\omega) - \gamma^2|\epsilon|^4},\\
a_\mathrm{s}(\omega) = \frac{\sqrt{\kappa_\mathrm{s, \mathrm{ex}}}(\kappa_\mathrm{i}/2 - i\omega) a_\mathrm{s, \mathrm{in}} + \sqrt{\kappa_\mathrm{s, 0}}(\kappa_\mathrm{i}/2 - i\omega) a_\mathrm{s, \mathrm{loss}} + \gamma\epsilon^2\sqrt{\kappa_\mathrm{i, \mathrm{ex}}} a_\mathrm{i, \mathrm{in}}^\dagger + \gamma\epsilon^2\sqrt{\kappa_\mathrm{i, \mathrm{0}}} a_\mathrm{i, \mathrm{loss}}^\dagger}{(\kappa_\mathrm{i}/2-i\omega)(\kappa_\mathrm{s}/2-i\omega) - \gamma^2|\epsilon|^4}.
\end{align}
\label{eq:result}
\end{subequations}

For SFWM, the pump power is well below the parametric oscillation threshold, thus $\gamma^2|\epsilon|^4$ in Eq. \ref{eq:result} can be dropped out. 
Note that the input of signal and idler modes are vacuum, thus the spectral density can be directly calculated
\begin{subequations}
\begin{align}
S_\mathrm{i}^\mathrm{out}(\omega) &= \langle a_\mathrm{i, \mathrm{out}}^\dagger a_\mathrm{i, \mathrm{out}}\rangle (\omega) = \frac{16\kappa_\mathrm{i, \mathrm{ex}}\kappa_\mathrm{s} \gamma^2|\epsilon|^4}{(\kappa_\mathrm{i}^2+4\omega^2)(\kappa_\mathrm{s}^2+4\omega^2)}, \\
S_\mathrm{s}^\mathrm{out}(\omega) &= \langle a_\mathrm{s, \mathrm{out}}^\dagger a_\mathrm{s, \mathrm{out}}\rangle (\omega) = \frac{16\kappa_\mathrm{s, \mathrm{ex}}\kappa_\mathrm{i} \gamma^2|\epsilon|^4}{(\kappa_\mathrm{i}^2+4\omega^2)(\kappa_\mathrm{s}^2+4\omega^2)}.
\label{eq:spectral}
\end{align}
\end{subequations}
From Eq. (\ref{eq:spectral}), one can see that the signal and idler photons have the same linewidth $\delta\nu$, which can be calculated as
\begin{equation}
  \delta\nu = \frac{1}{\sqrt{2}}\sqrt{\sqrt{(\kappa_\mathrm{i}^2+\kappa_\mathrm{s}^2)^2 + 4k_\mathrm{i}^2k_\mathrm{s}^2}-(\kappa_\mathrm{i}^2+\kappa_\mathrm{s}^2)}. 
  \label{eqn:linewidth}
\end{equation}
In our experiment, $\kappa_\mathrm{i}/2\pi=38.8$ MHz and $\kappa_\mathrm{s}/2\pi=41.8$ MHz result in $\delta\nu/2\pi=25.9$ MHz, which is used for the brightness calculation.

\clearpage
\section{Two-photon interference with the Franson interferometer}
\vspace{0.5cm}

\begin{figure*}[h!]
\centering
\includegraphics[width=0.45\textwidth]{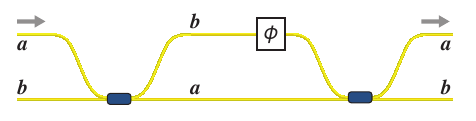}
\caption{
\textbf{A schematic of the Franson interferometer.} 
}
\label{Fig:S4}
\end{figure*}

The entanglement nature of photon pairs generated through cavity-enhanced SFWM process are post-selected by a Franson interferometer, i.e. an unbalanced Mach-Zehnder interferometer, as shown in Fig. \ref{Fig:S4}.
The interferometer's arm difference (corresponding to a time delay of $\Delta T$) determines the energy-time entangled state to be measured, written as
\begin{equation}
    |\Psi\rangle = 
    \frac{1}{\sqrt{2}}\left[a^\dagger_\mathrm{s}(t-\Delta T)a^\dagger_\mathrm{i}(t-\Delta T) + a^\dagger_\mathrm{s}(t)a^\dagger_\mathrm{i}(t)\right] 
    |0\rangle, 
\end{equation}
where $a^\dagger_\mathrm{s/i}$ is the creation operator for the path mode $a$. 
The temporal parameter $t$ denotes the time that the SNSPD clicks upon photon arrival. 
The two photons along the same path are then directed to the upper branch of the Franson interferometer, as shown in Fig. \ref{Fig:S4}. 
After the first beam splitter, the photon state is transformed to 
\begin{align*}
    |\Psi\rangle &\rightarrow \mathcal{N}\Bigg(\Big[a_\mathrm{i}^\dagger(t-\Delta T)+ib_\mathrm{i}^\dagger(t-\Delta T)\Big] \Big[a_\mathrm{s}^\dagger(t-\Delta T)+ib_\mathrm{s}^\dagger(t-\Delta T)\Big] + \Big[a_\mathrm{i}^\dagger(t)+ib_\mathrm{i}^\dagger(t)\Big] \Big[a_\mathrm{s}^\dagger(t)+ib_\mathrm{s}^\dagger(t)\Big]\Bigg)|0\rangle, 
\end{align*}
where $\mathcal{N}$ is a normalization coefficient. 
Then the branch $b$ is directed to the fiber link to acquire an additional delay of $\Delta T$. 
After passing a phase shifter introducing $\phi$, the branch $b$ is finally recombined with the branch $a$. 
The phase shifter, together with the second beam splitter, take the transformation $a^\dagger\rightarrow a^\dagger/\sqrt{2}$ and $b^\dagger \rightarrow ie^{i\phi} a^\dagger/\sqrt{2}$. 
Thus the state is further transformed to
\begin{align*}
    |\Psi\rangle &\rightarrow \mathcal{N} \Bigg(\Big[a_\mathrm{i}^\dagger(t-\Delta T)-e^{i\phi}a_\mathrm{i}^\dagger(t)\Big] \Big[a_\mathrm{s}^\dagger(t-\Delta T)-e^{i\phi}a_\mathrm{s}^\dagger(t)\Big] + \Big[a_\mathrm{i}^\dagger(t)-e^{i\phi}a_\mathrm{i}^\dagger(t+\Delta T)\Big] \Big[a_\mathrm{s}^\dagger(t)-e^{i\phi}a_\mathrm{s}^\dagger(t+\Delta T)\Big]\Bigg)|0\rangle.\\
    &= \mathcal{N}
    \Bigg(\left[a_\mathrm{i}^\dagger(t-\Delta T)a_\mathrm{s}^\dagger(t-\Delta T)+a_\mathrm{i}^\dagger(t)a_\mathrm{s}^\dagger(t)+e^{i2\phi}a_\mathrm{i}^\dagger(t)a_\mathrm{s}^\dagger(t)+e^{i2\phi}a_\mathrm{i}^\dagger(t+\Delta T)a_\mathrm{s}^\dagger(t+\Delta T)\right] \\
    &~~~~~-e^{i\phi}\left[a_\mathrm{s}^\dagger(t)a_\mathrm{i}^\dagger(t-\Delta T)+a_\mathrm{s}^\dagger(t-\Delta T)a_\mathrm{i}^\dagger(t)+a_\mathrm{s}^\dagger(t)a_\mathrm{i}^\dagger(t+\Delta T)+a_\mathrm{s}^\dagger(t)a_\mathrm{i}^\dagger(t+\Delta T)\right]\Bigg)|0\rangle. 
\end{align*}
The SNSPD cannot distinguish the events between $a_\mathrm{s}^\dagger(t)a_\mathrm{i}^\dagger(t-\Delta T)$ and $a_\mathrm{s}^\dagger(t+\Delta T)a_\mathrm{i}^\dagger(t)$. 
In other words, the two-photon events with the same temporal difference are indistinguishable. 
Thus the photon state at the end of the Franson interferometer can be written as
\begin{align}
    |\Psi\rangle 
    \rightarrow \mathcal{N}\Big[(1+e^{i2\phi})a_\mathrm{i}^\dagger(t)a_\mathrm{s}^\dagger(t)
    -e^{i\phi} a_\mathrm{i}^\dagger(t)a_\mathrm{s}^\dagger(t-\Delta T)
    -e^{i\phi} a_\mathrm{i}^\dagger(t)a_\mathrm{s}^\dagger(t+\Delta T)\Big]|0\rangle.
    \label{eqn:franson}
\end{align}
In Eq. (\ref{eqn:franson}), the first term corresponds to the central peak of the histogram, showing the quantum interference. 
While the second and the third terms correspond to the sidebands, whose height is independent of the phase $\phi$. 

\clearpage
\section{A comparison of integrated microresonator-based quantum light sources}
\vspace{0.5cm}

A comparison of integrated microresonator-based quantum light sources is shown in Tab. \ref{tab:2}. 
For publications that explicitly report the photon-pair generation rate, the brightness is estimated using the linewidth calculated by either Eq. (\ref{eqn:linewidth}) or $0.64\kappa/2\pi$ of the generated photons' mode. 
Our work set record values for brightness and linewidth for microresonator-based photon-pair sources demonstrated on CMOS-compatible platforms. 

\begin{table*}[h!]
\caption{\label{tab:table2}
A comparison of integrated microresonator-based photon-pair sources. 
The photon-pair generation rate (PGR)-, microresonator loaded $Q$ factor, photon linewidth, brightness, and two-photon interference visibility are listed. 
}
\begin{ruledtabular}
\begin{tabular}{l@{}cccccccc}
\multirow{2}*{Reference} & \multirow{2}*{Material} & CMOS & \multirow{2}*{Type} &PGR@1 mW & \multirow{2}*{Loaded $Q$} & Bandwidth & Brightness@1 mW & \multirow{2}*{Visibility} \\
 &  & compatible &  & (Hz) &  & (GHz) & (Hz/GHz) &  \\[1pt]
\colrule\\[-6pt]
\textbf{Our work} & \textbf{Si}$\bm{_3}$\textbf{N}$\bm{_4}$ & \textbf{Yes} & \textbf{SFWM} & $\bm{3.04\times10^7}$ & $\bm{5.0\times10^6}$ & \textbf{0.026} & $\bm{1.2\times10^9}$ & \textbf{97.3\%} \\
Fan \emph{et al}. \cite{Fan:23} & Si$_3$N$_4$ & Yes & SFWM & $1.15\times10^6$ & $1.0\times10^6$ & 0.12 & $9.6\times10^6$ & 97.3\% \\
Ramelow \emph{et al}. \cite{Ramelow:15} & Si$_3$N$_4$ & Yes & SFWM & $3.9\times10^6$ & $2\times10^6$ & 0.06 & $6.5\times10^7$ & 90\% \\
Lu \emph{et al}. \cite{Lu:19a} & Si$_3$N$_4$ & Yes & SFWM & $2.3\times10^6$ & $1.0\times10^6$ & 0.91 & $1.1\times10^7$ & 82.7\% \\
Ma \emph{et al}. \cite{Ma:17} & Silicon & Yes & SFWM & $1.5\times10^8$ & $9.2\times10^4$ & 2.1 & $1.6\times10^8$ & 95.9\% \\
Li \emph{et al.} \cite{LiJ:24} & 3C-SiC & Yes & SPDC & 9.6$\times10^{5}$ & 2.5$\times10^4$ & 5.0 & $1.9\times10^{5}$ & $86.0$\%\\[2pt]
Guo \emph{et al.} \cite{Guo:17a} & AlN & Yes & SPDC & 5.8$\times10^{6}$ & 2$\times10^5$ & 0.70 & $8.3\times10^{6}$ & $-$\\
\colrule\\[-6pt]
Steiner \emph{et al.} \cite{Steiner:21} & AlGaAs & No & SFWM & 2$\times10^{10}$ & 1.2$\times10^6$ & 0.10 & $2\times10^{11}$ & 97.1\%\\
Kummar \emph{et al.} \cite{Kumar:19} & InP & No & SFWM & 1.45$\times10^{8}$ & 4.3$\times10^4$ & 3.1 & $4.7\times10^{7}$ & $-$\\
Zeng \emph{et al.} \cite{ZengH:24} & GaN & No & SFWM & 2.09$\times10^{6}$ & 4.3$\times10^5$ & 0.3 & $7.0\times10^{6}$ & 95.5\%\\
Ma \emph{et al.} \cite{Ma:20} & PPLN & No & SPDC & 2.7$\times10^{9}$ & 1.0$\times10^5$ & 1.2 & $2.3\times10^{9}$ & $-$\\
\end{tabular}
\end{ruledtabular}
\label{tab:2}
\end{table*}

\section*{Supplementary References}
\bibliographystyle{apsrev4-1}
%